\documentclass[aps,prd,onecolumn,groupedaddress,showpacs,nofootinbib,amssymb]{revtex4}
\usepackage[dvips]{graphicx}
\usepackage{amssymb}
\usepackage{amsmath}
\usepackage{graphicx,,color}
\usepackage{amsfonts}
\usepackage{bm}
\usepackage{comment}
\usepackage{color}

\begin{document}

\newcommand{\be}{\begin{equation}}
\newcommand{\ee}{\end{equation}}
\newcommand{\bea}{\begin{eqnarray}}
\newcommand{\eea}{\end{eqnarray}}
\newcommand{\nn}{\nonumber \\}
\newcommand{\e}{\mathrm{e}}

\title{Dark Energy and Cosmological Horizon Thermal Effects}
\author{Artyom~V.~Astashenok$^{1}$,
Sergei~D.~Odintsov$^{2,3}$,V.K. Oikonomou$^{4,5}$}
\affiliation{$^1$ I. Kant Baltic Federal University, Institute of Physics, Mathematics and IT, 236041, 14, Nevsky st., Kaliningrad, Russia \\
$^2$Instituci\`{o} Catalana de Recerca i Estudis Avan\c{c}ats
(ICREA),
Barcelona, Spain \\
$^3$Institut de Ciencies de l'Espai (CSIC-IEEC), Campus UAB,
Facultat de Ciencies, Torre C5-Par-2a pl, E-08193 Bellaterra
(Barcelona), Spain\\
$^{4}$ Department of Physics, Aristotle University of
Thessaloniki, Thessaloniki 54124,
Greece\\
$^{5}$ Laboratory for Theoretical Cosmology, Tomsk State
University of Control Systems and Radioelectronics, 634050 Tomsk,
Russia (TUSUR)}

\begin{abstract}
We investigate various dark energy models by taking into account
the thermal effects induced from Hawking radiation on the apparent
horizon of the Universe, for example near a finite-time future
singularity. If the dark energy density increases as the Universe
expands, the Universe's evolution reaches a singularity of II type
(or sudden future singularity). The second derivative of scale
factor diverges but the first remains finite. Quasi-de Sitter
evolution can change on sudden future singularity in the case of
having an effective cosmological constant larger than the maximum
possible value of the energy density of the Universe. Another
interesting feature of cosmological solution is the possibility of
a transition between deceleration and acceleration for
quintessence dark energy with a simple equation of state. Finally,
we investigate which fluid component can remedy Big Rip
singularities and other crushing type singularities.
\end{abstract}

\maketitle

\section{Introduction}

The accelerated expansion of the Universe \cite{Riess,Perlmutter}
from a theoretical viewpoint can be caused by some fluid with
negative pressure and/or negative entropy (for review, see
\cite{Bamba:2012cp,Dark-6,Cai:2009zp}). Due to the fact that it is
still unknown how this fluid will behave, the late-time
accelerated expansion era is dubbed the dark energy era. The
current observations coming from Planck \cite{Akrami:2018odb},
indicate that dark energy controls nearly 70\% of total energy
density of the Universe \cite{Kowalski}. The equation of state
(EoS) parameter for the dark fluid, namely $w_{d}$, is negative,
that is,
 \be
w_\mathrm{d}=p_{d}/\rho_{d}<0, \ee where $\rho_{d}$ and $p_{d}$
are the dark energy density and pressure correspondingly. However,
it is still not clear what is the precise value of $w_d$
\cite{PDP,Amman}, although the latest Planck data constraint
significantly the values that the EoS parameter can take. The
standard $\Lambda$-Cold-Dark-Matter ($\Lambda$CDM) model according
to which dark energy is simply a cosmological constant, implies
that $w_d=-1$. If the EoS parameter is in the range
$-1<w_{d}<-1/3$ the evolution of the Universe is quintessential.
An interesting case (although controversial \cite{Carrol}) arises
for $w_d<-1$, and in this case the Universe evolves dominated by
phantom dark energy. A simple choice of the form
$w_d=\mbox{const}$ leads to a Big-Rip singularity
\cite{Caldwell,Caldwell-2003,Carrol,P,Frampton,Nojiri:2003vn,Faraoni:2001tq,GonzalezDiaz:2003rf,Elizalde:2004mq,Singh:2003vx,Csaki:2004ha,Wu:2004ex,Nesseris:2004uj,Stefancic:2003rc,Chimento:2003qy,Hao:2004ky,Dabrowski:2004hx,Nojiri}, where the scale factor and all the physical quantities that can be
defined on the three dimensional spacelike singularity defined of
the time instance that the Rip occurs, severely diverge. There are
other types of singularities for phantom Universe, for example a
singularity of type III corresponds to a situation in which, the
Hubble parameter $H$ diverges at a finite time and for a finite
scale factor. A sudden future singularity (or type II) is milder
in comparison with the two previous. The scale factor and its
first derivative (and therefore the energy density) are finite at
the moment of the singularity, but the second derivative of scale
factor diverges (and the pressure correspondingly)
\cite{Barrow,Shtanov:2002ek,Nojiri:2004ip,Cotsakis:2004ih,Dabrowski:2004bz,FernandezJambrina:2004yy,FernandezJambrina:2008dt,Barrow:2004he,Stefancic:2004kb,Cattoen:2005dx,Tretyakov:2005en,Balcerzak:2006ac,Yurov:2007tw,Barrow:2009df,BouhmadiLopez:2009jk}. This singularity can occur for quintessential
evolution too. In principle, the occurrence of finite-time
singularities, during the evolution of the Universe, is not
necessarily a consequence of phantom dark energy. In fact, a
phantom scalar field always leads to a Big Rip singularity
\cite{Caldwell}, but the opposite is not true. For some studies on
alternative models leading to a little-rip singularity, see Refs.
\cite{Frampton-2,Frampton:2011aa,Frampton-3,Astashenok}. If the EoS parameter $w$
tends to minus unity sufficiently fast, the time instance that the
singularity occurs is at infinity. Of course, by that time, the
tidal forces  would grow infinitely large and all the stellar
structures and galactic structures would be torn apart by the
tidal forces before the singularity would ever be reached. Another
possibility is the so-called pseudo-rip scenario, which can be
realized when the energy density of dark energy approaches
asymptotically a constant value (effective ``cosmological
constant''). This effective ``cosmological constant'' in principle
can considerably differ from the current energy density of dark
energy $\rho_{d0}$.

{According to the observational data, it is marginally probable
that the phantom divide line is crossed and the dark energy era is
actually a phantom dark energy era. For example for the flat
$wCDM$ model according to latest Planck data, the dark energy EoS
is constrained as follows,
$$
w=-0.978\pm 0.059\, ,
$$
so the possibility $w<-1$ is not excluded, and the dark energy era
is a probable scenario. But for $w<-1$ future finite time
singularities can occur. Therefore, the occurrence of future
finite time singularities is a realistic possibility, even present
in the context of Einstein-Hilbert gravity by using phantom scalar
fields \cite{Caldwell-2003}.} When a finite-time singularity is
approached, it is natural to assume that several effects having to
do with the cosmological horizon surrounding the singularity,
should be taken into account. The crushing-type finite-time
singularities are somewhat similar to black hole singularities, in
the sense that these are both future spacelike singularities. In
this sense, it is natural to assume that the cosmological horizon
would cause some cosmological effects on the evolution of the
Universe, near the singularity, exactly as in the black holes
case. Particularly, using the analogy with Hawking radiation from
black holes one can consider thermal radiation from horizon of the
Universe \cite{Gibbons}, \cite{Cai}. In \cite{Rugg} it was argued
that the kind of cyclic cosmology might be realized instead of the
Big Rip singularity due to effect of thermal radiation. In this
paper we shall consider the effects of thermal cosmological
horizon terms, generalizing the work of \cite{Nojiri:2020sti}, and
for the dark energy we shall assume that it will have a
generalized EoS of the form $p_d=g(\rho_{d})$, where $g$ is an
arbitrary function. We shall consider various models of dark
energy with the inclusion of thermal radiation effects. After
discussing the main theoretical framework, we shall study two
important state-finder parameters, namely the deceleration and
jerk parameters. Then some simple dark energy EoS parameters are
considered, with account of thermal radiation. Also our findings
show that for any equation of state which leads to infinitely
increasing phantom energy density, a sudden future singularity
always occurs. Also the little rip regime cannot be realized in
the present framework, and interestingly enough only one scenario
of non-singular evolution for phantom energy takes place, that of
pseudo-rip evolution, but in that case, the value of the
''effective'' cosmological constant should be less than some
critical density. It is also important to note that in the present
context, a phantom dark energy era can be realized without the
need for a phantom scalar field, as in ordinary Einstein-Hilbert
gravity.

\section{Field Equations and the Description of the Theoretical Framework \label{SecII}}

In this section we briefly consider the cosmological equations for
the evolution of the Universe, by taking into account the thermal
radiation effects from the cosmological horizon. Let's start from
the spatially-flat FRW universe with metric, \be
ds^{2}=dt^{2}-a^{2}(t)(dx^{2}+dy^{2}+dz^{2}). \ee The cosmological
equations are, \be \label{Fried1} \left(\frac{\dot a}{a}\right)^2
= \frac{\rho}{3}\, , \quad \frac{\ddot{a}}{a} = -\frac{1}{6}(\rho
+ 3 p)\, , \ee where $\rho$ and $p$ are the total energy density
and pressure, $a$ is the scale factor,and the ``dot'' denotes
differentiation with respect to the cosmic time. We use the system
of units in which $8\pi G=c=1$, the so-called reduced Planck units
system.

From the Friedmann and the Raychaudhuri equations it follows that,
\begin{equation}\label{conserv}
\dot{\rho}+3H(\rho+p)=0.
\end{equation}
If the Universe consists of non-interacting components $\rho_{i}$,
this equation holds for each component $\rho_{i}$.

The radius of the apparent horizon of the Universe is
$$
r_h\sim 1/H\, ,
$$
Therefore, the Hawking temperature of the thermal radiation
appearing due to the apparent horizon is proportional to Hubble
parameter $H$. The energy density of radiation thermal term
according to Stefan-Boltzmann law is,
$$
\rho_{t_{rad}}=3\alpha H^4\, ,
$$
where $\alpha$ is some constant. The multiplicative factor 3 is
introduced for convenience. {One can expect that the constant
$\alpha$ could be calculated from a self-consistent quantum model
describing the thermal radiation. Such a model though is not
available, and therefore the direct usage of the thermal radiation
effects related to the de Sitter horizon is not justified.
However, in order to have a concrete idea of these effects, we
assumed only by analogy that the energy density of thermal
radiation is proportional to $H^4$, because the temperature of
Hawking radiation of the apparent horizon is inversely
proportional to the size of the horizon.

A similar physical analogy is used in the context of holographic
energy, for which we usually assume that,
$$
\rho = \beta /L^2\, ,
$$
where $L$ is the event horizon. Regarding the values of $\beta$,
there are various estimations from calculations, but its exact
value is unknown. Therefore we use $\alpha$ as the unknown
parameter in our calculations.}

The Friedmann equation for the Hubble parameter, by taking into
account the density of thermal radiation, is,
\begin{equation}
\alpha H^4 - H^2 + \frac{\rho}{3} = 0\, ,
\end{equation}
which can be solved with respect to the Hubble parameter as
follows,
\begin{equation}\label{hubble}
H = \pm \left(\frac{1\pm
\sqrt{1-4\alpha\rho/3}}{2\alpha}\right)^{1/2}\, .
\end{equation}
Further we assume that the Universe expands, and therefore the
sign ``+'' is chosen. One also needs to choose sign ``-'' in
brackets for the following reason. {Seing the thermal radiation as
the Hawking radiation from the apparent horizon has a quantum
origin. It is known that the cosmological evolution of Universe
from beginning to present date, is described well by the Friedmann
equations and one can expect that for $\alpha\rightarrow 0$
(Universe without thermal radiation) we should obtain the ordinary
Friedmann equations.} And for $\alpha\rho/3<<1$, we obtain the
ordinary cosmological model,
$$
H^2=\frac{1-\sqrt{1-4\alpha\rho/3}}{2\alpha}\approx \frac{1-(1-2\alpha\rho/3)}{2\alpha}=\frac{\rho}{3}.
$$
But by choosing ``+'', we obtained very strange cosmological model
with {a Hubble parameter} which tends to infinity for
$\alpha\rightarrow\infty$.

For these choices of signs in Eq. (\ref{hubble}), one can obtain
the following relation for the derivative of Hubble parameter,
\begin{equation}
\dot{H}=-\frac{1}{2}\frac{\rho+p}{\sqrt{1-4\alpha\rho/3}}.
\end{equation}
The second derivative of the scale factor is,
\begin{equation}
\frac{\ddot{a}}{a}=\dot{H}+H^2=-\frac{1}{2}\frac{\rho+p}{\sqrt{1-4\alpha\rho/3}}+\frac{1-
\sqrt{1-4\alpha\rho/3}}{2\alpha}.
\end{equation}
We consider the case when equation of state for dark energy is
given in form,
\begin{equation} \label{EoS}
p_\mathrm{d}=-\rho_\mathrm{d}-f(\rho_\mathrm{d})\, ,
\end{equation}
where $f(\rho_\mathrm{d})$ is some arbitrary function. Condition
$f(\rho)>0$ corresponds to phantom energy while for $f(\rho)<0$,
describes the case of quintessential evolution.

For the case of dominating of dark energy we can don't consider
other components of matter. In this case from the equation for
Hubble parameter and Eq. (\ref{conserv}), one can get the relation
for cosmological time,
\begin{equation}\label{trho}
t = \frac{\sqrt{2\alpha}}{{3}}\int^{\rho}_{\rho_{0}} \frac{d
\rho}{f(\rho)\left(1-\sqrt{1-4\alpha\rho/3}\right)^{1/2}}.
\end{equation}
We omit the subscript $\mathrm{d}$ in this relation assuming $\rho=\rho_{d}$.
The quintessence energy density decreases with time
($\rho<\rho_{0}$), while the phantom energy density increases
($\rho>\rho_{0}$).

The effective pressure and density can be defined as,
\begin{equation}
\rho_{eff}=3H^2, \quad p_{eff}=-2\dot{H}-3H^2
\end{equation}
and therefore for our cosmological model we have,
\begin{equation}
\rho_{eff}=\frac{3}{2\alpha}\left(1-\sqrt{1-4\alpha\rho/3}\right),
\end{equation}
$$
p_{eff}=\frac{\rho+p}{\sqrt{1-4\alpha\rho/3}}-\rho_{eff}.
$$
From this relation one can conclude that the effective pressure
reaches infinity for a finite value of the energy density
$\rho_{max}=\frac{3}{4\alpha}$ (the effective energy density is
finite also). Therefore, we have a type II singularity (or sudden
future singularity in another words). We can also compute the
scale factor as,
\begin{equation}\label{at}
a=a_{0}\exp\left(\frac{1}{3}\int^{\rho}_{\rho_{0}}\frac{d\rho}{f(\rho)}\right)\,
.
\end{equation}
From Eq. (\ref{trho}) one can see that there are two main
possibilities for the cosmological evolution,
\begin{enumerate}
\item The integral (\ref{trho}) converges at
$\rho\rightarrow\rho_{m}=\frac{3}{4\alpha}$ and singularity of
type II takes place \cite{Barrow,Nojiri-2}.

The scale factor remains finite with its first derivative with
respect to the cosmic time. This variant of cosmological evolution
is realized for phantom dark energy only. Also we have singularity
of this type if $f(\rho)$ has a zero at some $\rho_{f}<\rho_m$ and
the integral with respect to the cosmic time converges.

\item {Although we do not know the equation of state for dark
energy, we can forecast its qualitative behavior by assuming that
the function $f(\rho)$ has a zero at some value $\rho_f$.} If the
integral (\ref{trho}) diverges at some value of energy density
$\rho\rightarrow\rho_{f}<\frac{3}{4\alpha}$. In this case scale
factor tends to infinity at $t\rightarrow\infty$. Therefore we
have quasi-de Sitter expansion with effective cosmological
constant $\Lambda=\rho_{f}$.
\end{enumerate}

{Of course one can ask how it is possible to discriminate the
effects of the thermal radiation from modified gravity or specific
fluid effects. This question arises for many cosmological models.
Theories of modified gravity can lead to cosmological dynamics
similar to models with some fluids or models on brane for example.
Maybe one of the arguments is in favor for the model of thermal
radiation which is inherently relatively simpler (on our opinion
of course) in comparison with complicated form of actions of
Galileon gravity for example.}

One of the main indicators for any cosmological model is data
about distance modulus as a function of the redshift from the
Supernova Cosmology Project \cite{Amanullah:2010vv}. For standard
cosmology the distance modulus for a supernova with redshift
$z=a_{0}/a-1$ is \be \mu(z)=\mbox{const}+5\log D_{L}(z). \ee Here
$D_{L}(z)$ is the luminosity distance. For Friedmann cosmology,
the  luminosity distance is,
\begin{equation}
    D_{L}(z)=\frac{c}{H_{0}}(1+z)\int_{0}^{z}\frac{dz}{E(z)},
\end{equation}
where $E(z)$ is the dimensionless Hubble parameter i.e.
$$
E(z)=\frac{H(z)}{H_{0}}=\left(\frac{\rho}{\rho_{0}}\right)^{1/2}.
$$
In particular, for the well known $\Lambda$CDM model, we have, \be
\label{DLSC} E(z)=
\left(\Omega_{m}(1+z)^{3}+\Omega_{\Lambda}\right)^{1/2}. \ee Here,
$\Omega_{m}$ is the fraction of the total density contributed by
matter, and $\Omega_{\Lambda}$ is the fraction contributed by the
vacuum energy density.

For our purposes, it is convenient to use dimensionless units for
the Hubble parameter, $\alpha$ and energy density:
$$
\alpha\rightarrow \tilde{\alpha}H_{0}^{-2},\quad H\rightarrow H_{0}E,
\quad \rho \rightarrow 3H_{0}^2\tilde{\rho}.
$$
In these units for dimensionless Hubble parameter, we obtain,
\begin{equation}
    E=\left(\frac{1-\sqrt{1-4\tilde{\alpha}\tilde{\rho}}}{2\tilde{\alpha}}\right)^{1/2}.
\end{equation}
One needs to take into account that, at present time, the
dimensionless density satisfies $\tilde{\rho}_{0}\neq 1$, in
contrast with the ordinary  Friedmann cosmological model, because
$\rho_{0}\neq 3H_{0}^2$ although for $\tilde{\alpha}<<1$ one can
assume that $\tilde{\rho}_{0}\approx 1$.

The deceleration parameter $q_{0}$ is defined according to
relation,
\begin{equation}
q=- \frac{1}{a H^2} \frac{d^2 a}{dt^2}=
\end{equation}
$$
=-1+{\alpha}\left(1-4\alpha\rho/3\right)^{-1/2}\left(1-(1-4\alpha\rho/3)^{1/2}\right)^{-1}(\rho+p)
$$
In Friedmann cosmology one obtains the well-known expression for
$q$ assuming simply $\alpha\rightarrow 0$ in the following
equation,
\begin{equation}
    q^{(0)}=-1+\frac{3}{2}\left(1+\frac{p}{\rho}\right).
\end{equation}
For the $\Lambda$CDM model one obtains,
$$
q^{(0)}_{\Lambda CDM}=-1+\frac{3}{2}\Omega_{m}.
$$
One can get also jerk parameter $j$ \cite{Sahni:2002fz} for our
model in comparison with ordinary cosmology,
\begin{equation}
\label{DDD1}
j = \frac{1}{a H^3} \frac{d^3 a}{dt^3}=
\end{equation}
$$
=2\alpha\left(\frac{\alpha(\rho+p)^2}{(1-4\alpha\rho/3)^{3/2}}+\frac{1-\sqrt{1-4\alpha\rho/3}}{2\alpha}+\frac{3(\rho+p)}{2(1-4\alpha\rho/3)^{1/2}}\frac{dp}{d\rho}\right)(1-(1-4\alpha\rho/3)^{1/2})^{-1}
$$
For $\alpha\rightarrow 0$ we get,
\begin{equation}
    j^{(0)}=1+\frac{9}{2}\left(1+\frac{p}{\rho}\right)\frac{dp}{d\rho}
\end{equation}
and particularly for the $\Lambda$CDM model we have
$p=-\Lambda=\mbox{const}$, therefore $j^{(0)}_{\Lambda CDM}=1$.

Data about the deceleration and jerk parameters can be obtained
from astronomical observations of distant objects at redshifts
$z\gtrsim 1$. {Estimations for deceleration and jerk parameter
from current observations do not discriminate directly the correct
model of dark energy, because statistical errors are sufficiently
large, and therefore models alternative to the $\Lambda$CDM model
can be perfectly compatible to the current observational data.}
The $\Lambda$CDM values of these statefinder quantities can act as
reference points for alternative to $\Lambda$CDM models.

In the next sections, we shall present several dark energy models
by taking into account thermal radiation effects.

{There are various scenarios for a Universe with dark energy in
the future. From observations it is apparent that equation of
state for dark energy is very close to simple equation of state of
the cosmological constant,
$$
p_{d}=-\rho_{d}.
$$
It is convenient to use EoS formalism for other models of dark
energy. Such formalism can be applied not only to models in which
dark energy is a fluid but for scalar fields too. Our main purpose
was to investigate the role which thermal radiation in these
scenarios. Therefore the models that we shall study in the
following sections, are illustrative and simple examples of dark
energy models with a variety of possible future evolutions, with a
common characteristic though, that they mimic the standard
cosmological model in the cosmological past.}


\section{A Decelerating Quintessence Model}

It is well-known that for a quintessential Universe  with simple
equation of state $p=w\rho$ ($-1<w<-1/3$), only acceleration takes
place. By taking into account the thermal radiation effects,
allows us to construct models with a phase of deceleration in the
past, emerging from sudden finite-time singularities. Such
Universe emerges from a sudden finite-time singularity in the
past. The initial value of the Hubble parameter is,
\begin{equation}
    H_{i}=\frac{1}{\sqrt{2\alpha}},
\end{equation}
and the energy density of quintessence at the moment of the
singularity is,
\begin{equation}
    \rho_{i}=\frac{3}{4\alpha}
\end{equation}
and then it decreases with scale factor as,
$$
\rho_d=\frac{\rho_i a_{i}^{3(w+1)}}{a^{3(w+1)}}.
$$
Therefore we have,
\begin{equation}\label{accel}
    \frac{\ddot{a}}{a}=-\frac{1}{2}
    \frac{\rho_{i}(1+w)}{\sqrt{1-a_{i}^{3(w+1)}/a^{3(w+1)}}}
    \frac{a_{i}^{3(w+1)}}{a^{3(w+1)}}+\frac{1-\sqrt{1-a_{i}^{3(w+1)}/a^{3(w+1)}}}{2\alpha}\,
    .
\end{equation}
For $t>>t_i$ we have,
$$
\frac{\ddot{a}}{{a}}=-\frac{1}{6}(1+3w)\rho_i a_{i}^{3(w+1)}/a^{3(w+1)}>0.
$$
Near the initial singularity, the first term in Eq. (\ref{accel}) is
very large and negative for quintessence, and therefore
$\ddot{a}<0$. Eventually a deceleration to acceleration transition
occurs, at some later time instance. Of course this cosmological
model is not realistic, since we did not aim to confront it with
the observational data. It is a rather qualitative model, showing
that a decelerating quintessential model which evolves to an
accelerating state, is possible to be realized by taking into
account thermal effects.

{In Fig. 1 we depicted dependence of acceleration from
scale factor for various $w$ and $\alpha$.
Duration of decelerating phase depends from $w$ strongly. In a
phantom case first term is always positive and therefore universe
filled phantom energy expands with acceleration.}

\begin{figure}
    \centering
    \includegraphics[scale=0.35]{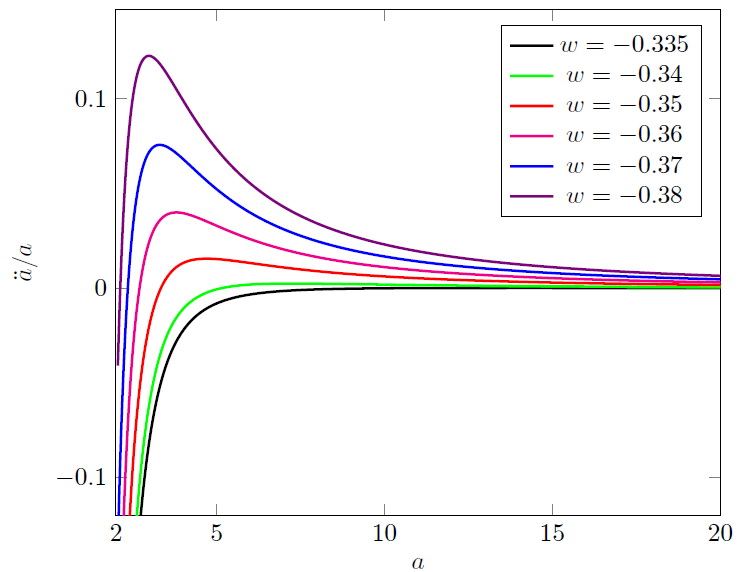}
    \caption{Dependence of acceleration $\ddot{a}/a$ (in units of $H_{0}^{2}$ from scale factor for various values of $w$ in a case of simple model of quintessence with constant
parameter of state. For $\tilde{\alpha}$ we take value 0.01.
Initial value of scalar factor is $a_i=1$. Duration of phase with
$\ddot{a}<0$ decreases for larger $|w|$. Maximal value of
acceleration otherwise increases.}
    \label{fig:accel}
\end{figure}


\section{Transition from Little-Rip and Type I and III Singularities to type II Singularity \label{SecIII}}

Let's consider the case of the following EoS,
\begin{equation}\label{EOS1}
f(\rho_{d})=\beta^2 \rho_{d0}(\rho_{d}/\rho_{d0})^{\gamma},\quad
\beta,\gamma\equiv \mbox{const}>0.
\end{equation}
In Friedmann cosmology without radiation term, this EoS leads to
various scenarios of cosmological evolution: 1) little rip takes
place for $\gamma\leq 1/2$ \cite{Frampton-2}-\cite{Astashenok}; 2)
$1/2<\gamma\leq 1$ corresponds to Big Rip singularity and 3) for
$\gamma>1$ we have a Type III singularity. The EoS parameter
$w=p_{d}/\rho_{d}$ for the above EoS at present time is simply,
$$
w_{0}=-1-\beta^2.
$$
In our case we have only Type II singularity for this EoS. We
start our consideration using a simple model: \be \label{LRT}
f(\rho)=\beta^{2}\rho_{d0}=\mathrm{const}\, . \ee From the
continuity equation for the dark energy fluid (\ref{Fried1}) we
derive \be\label{rhoLRT} \rho_{d}=\rho_{d0}(1+3\beta^{2}\ln a).
\ee Without loss of generality we assume that the current value of
the scale factor is $a_0=1$. We consider the Universe filled
matter and dark energy only and calculate the time for the final
singularity for several values of $w_{0}$ and $\alpha$. Also we
considered the behavior of cosmological acceleration in past and
we found the time instance at which $\ddot{a}$ changes sign, and
our results are presented in \ref{table1}.
\begin{table}\label{table1}
\begin{tabular}{|c|c|c|c|c|c|c|}
\hline
    & \multicolumn{2}{c}{$w_{0}=-1.05$}\vline &
\multicolumn{2}{c}{$w_{0}=-1.075$}\vline &
\multicolumn{2}{c}{$w_{0}=-1.10$}\vline \\
\hline
$\tilde{\alpha}$ & $t_\mathrm{f}-t_{0}$ & $t_{0}-t_{a}$ & $t_\mathrm{f}-t_{0}$ & $t_{0}-t_{a}$ &
$t_\mathrm{f}-t_{0}$ & $t_{0}-t_{a}$ \\
\hline
0.001 & 144.64 & 0.467 & 119.21 & 0.468 & 103.40 & 0.470 \\
0.005 & 107.69 & 0.460 & 71.77 & 0.462 & 53.82 & 0.464 \\
0.01 & 71.54 & 0.452 & 47.67 & 0.454 & 35.74 & 0.456 \\
0.02 & 45.99 & 0.435 & 30.64 & 0.438 & 22.97 & 0.440 \\
0.05 & 23.36 & 0.386 & 15.55 & 0.390 & 11.65 & 0.394 \\
\hline
\end{tabular}
\caption{Time for singularity $t_f-t_{0}$ and time after moment
when acceleration began $t_{0}-t_{a}$ (in units of $H_{0}^{-1}$)
for various values of $\tilde{\alpha}$). For initial $\Omega_m$
the value 0.28 is taken.}
\end{table}

\begin{figure}
    \centering
    \includegraphics[scale=0.28]{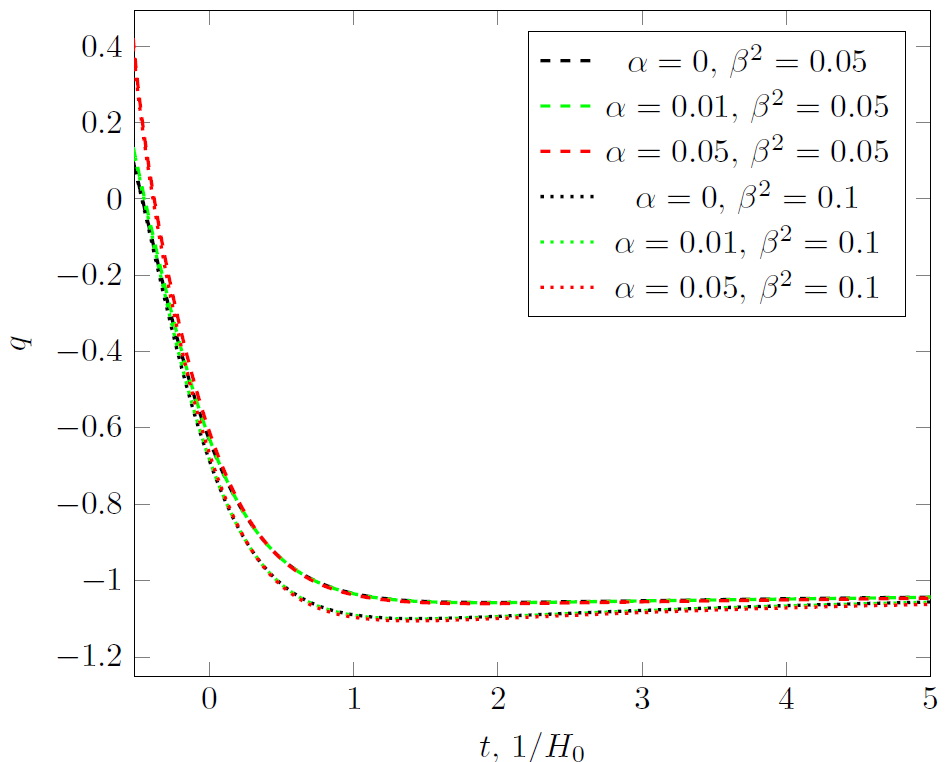}\includegraphics[scale=0.28]{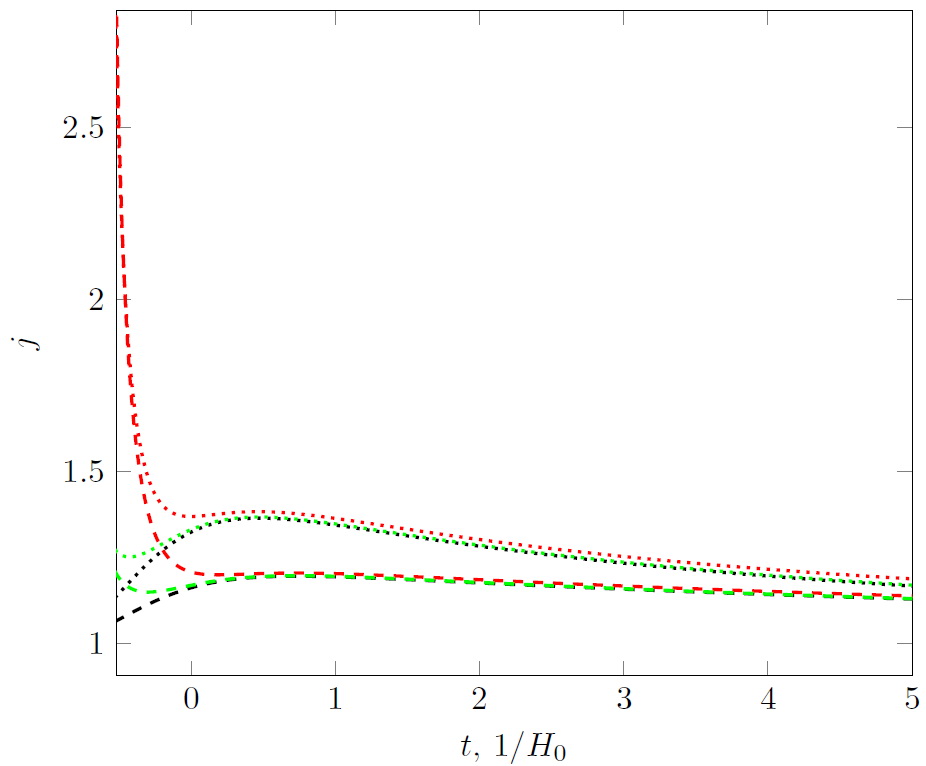}
    \caption{Deceleration (left panel) and jerk (right panel) parameters in
model ({\ref{LRT}}) for some $\beta^2$ and $\alpha$. Hereinafter
for $\Omega_{d0}$ is taken 0.72 in current moment.}
    \label{fig:1}
\end{figure}

The behavior of the deceleration $q$ and jerk $j$ parameters are
presented in Fig. \ref{fig:1} in comparison with dark energy model
(\ref{EOS1}) without thermal radiation from the moment of
beginning of cosmological acceleration.
Roughly, one can expect that thermal radiation effects have a
measurable effect, only during the cosmological acceleration.

One can describe this model in terms of scalar field theory, and
in the case of a phantom scalar field, the energy density and the
pressure of the scalar field are,
\begin{equation}
    \rho_d=-\frac{\dot{\phi}^2}{2}+V(\phi),\quad p_d = -\frac{\dot{\phi}^2}{2}-V(\phi).
\end{equation},
From these equations one can obtain that
\begin{equation}
    \phi=\int\sqrt{-\rho_d-p_d} dt= \int\sqrt{f(\rho_d)} dt.
\end{equation}
For $f(\rho)$ from Eq. (\ref{LRT}) we have simple linear
dependence of scalar field with respect to the cosmic time,
 \be
\phi=\phi_{0}+\beta \sqrt{\rho_{d0}} t. \ee For a Universe
dominated of dark energy without thermal radiation,
\begin{equation}
    t=\sqrt{3}\int\frac{da}{a\sqrt{\rho_{d}}}
\end{equation}
and thus we have for the integral (\ref{rhoLRT}),
$$
t=\frac{1}{\mu}\left({1+3\beta^2\ln a}-1\right),\quad \mu\equiv \frac{\beta^2\sqrt{3\rho_{d0}}}{2}.
$$
For the scalar field potential we have,
$$
2V=\rho_{d}-p_{d}=2\rho_d+f(\rho_{d})=2\rho_{d0}\sqrt{1+3\beta^2\ln a}+\beta^2\rho_{d0}.
$$
Using the expression for the cosmic time, one obtains the
potential as function of scalar field has the following form,
\begin{equation}
    V(\phi)=\frac{3\beta^2}{4}(\phi-\phi^{*})^{2}+\frac{\beta^2\rho_{d0}}{2},\quad \phi^*\equiv \phi_{0}-\frac{2}{\sqrt{3}\beta}
\end{equation}
With thermal radiation we have for the cosmic time a more
complicated expression,
\begin{equation}
    t=\frac{1}{\nu}\left\{\left(2+g(x)\right)\sqrt{1-g(x)}-\left(2+g(0)\right)\sqrt{1-g(0)}\right\}, \quad g(x)\equiv \sqrt{1-4\alpha\rho_{d0}/3-4\alpha\beta^2 x},
\end{equation}
$$
x\equiv\ln a,\quad \nu\equiv\frac{3\sqrt{\alpha}\beta^2\rho_{d0}}{\sqrt{2}}.
$$
Therefore, the potential of scalar field could be obtained only in
parametric form. {For comparison we present the
dependence of the potential with respect to various parameters in
Fig. 3.}

\begin{figure}
    \centering
    \includegraphics[scale=0.35]{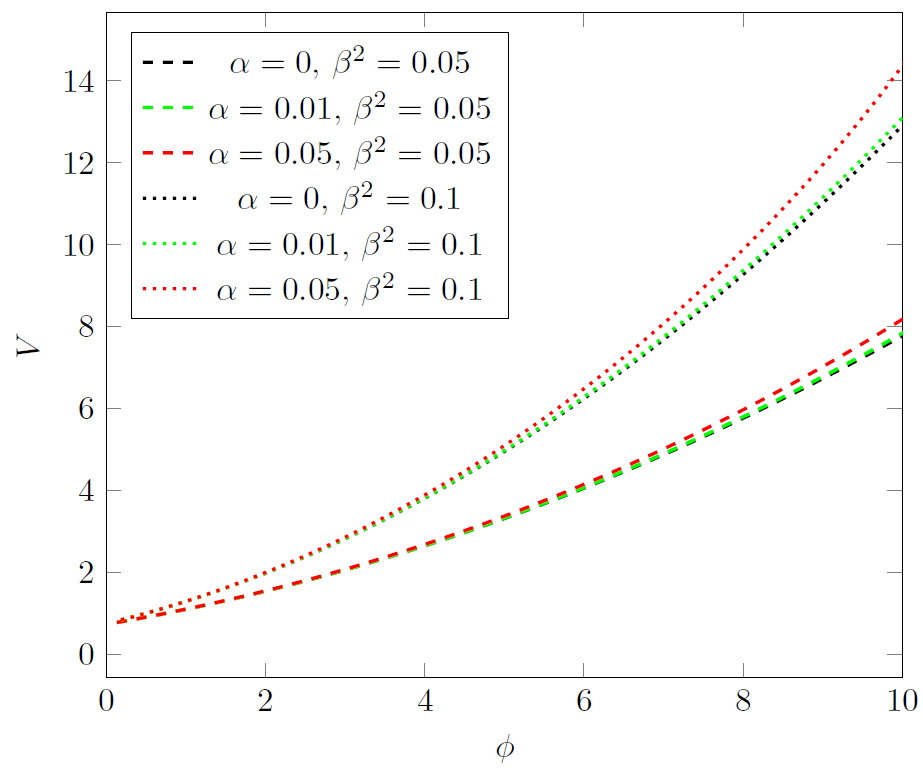}
    \caption{Potential of scalar field for model ({\ref{LRT}}) for the same parameters as on Fig. \ref{fig:1}.}
    \label{fig:V}
\end{figure}

For $\gamma=2$ we have the following dependence density of dark energy from scale factor
$$
\rho_{d}=\frac{\rho_{d0}}{1-3\beta^2\ln a}.
$$
The value of $a_{f}=\e^{\beta^{-2}/3}$ corresponds to the moment
of final singularity. The energy density tends to infinity for a
finite value of the scale factor. With account of thermal
radiation, the Universe ends its existence earlier not approaching
$a_f$. We compared the time for singularity occurrence in this
model without thermal radiation and with account of it, in Table
\ref{table2}.
\begin{table}\label{table2}
\begin{tabular}{|c|c|c|c|c|c|c|}
\hline
    & \multicolumn{2}{c}{$w_{0}=-1.05$}\vline &
\multicolumn{2}{c}{$w_{0}=-1.075$}\vline &
\multicolumn{2}{c}{$w_{0}=-1.10$}\vline \\
\hline
$\tilde{\alpha}$ & $t_{f}-t_{0}$ & $t_{0}-t_{a}$ & $t_{f}-t_{0}$ & $t_{0}-t_{a}$ &
$t_{f}-t_{0}$ & $t_{0}-t_{a}$ \\
\hline
0   & 5.176 &    0.469       & 3.432 &  0.469 & 2.562 & 0.469 \\
0.001 & 5.169 &   0.465      & 3.428 & 0.465  & 2.558 & 0.465   \\
0.005 & 5.140 &   0.459      & 3.408 &  0.459  & 2.544 & 0.459  \\
0.01 & 5.099 &    0.450      & 3.381 &  0.451  & 2.523 & 0.451\\
0.02 & 5.007 &     0.434     & 3.320 &  0.435  & 2.477 & 0.436 \\
0.05 & 4.680 &     0.384     & 3.101 &  0.387  & 2.312 & 0.389 \\
\hline
\end{tabular}
\caption{Time for singularity $t_f-t_{0}$ and time after moment
when acceleration began $t_{0}-t_{a}$ in model with thermal
radiation (in units of $H_{0}^{-1}$) for $\gamma=2$ in EoS
(\ref{EOS1}). For initial $\Omega_m$ the value 0.28 is taken.}
\end{table}
For the considered values of $\alpha$, the cosmological evolution
in the past does not differ significantly from the model without
thermal radiation. {The behavior of the parameters $q$
and $j$ for $\gamma=2$ can be seen in Fig. \ref{fig:2}.

\begin{figure}
    \centering
    \includegraphics[scale=0.28]{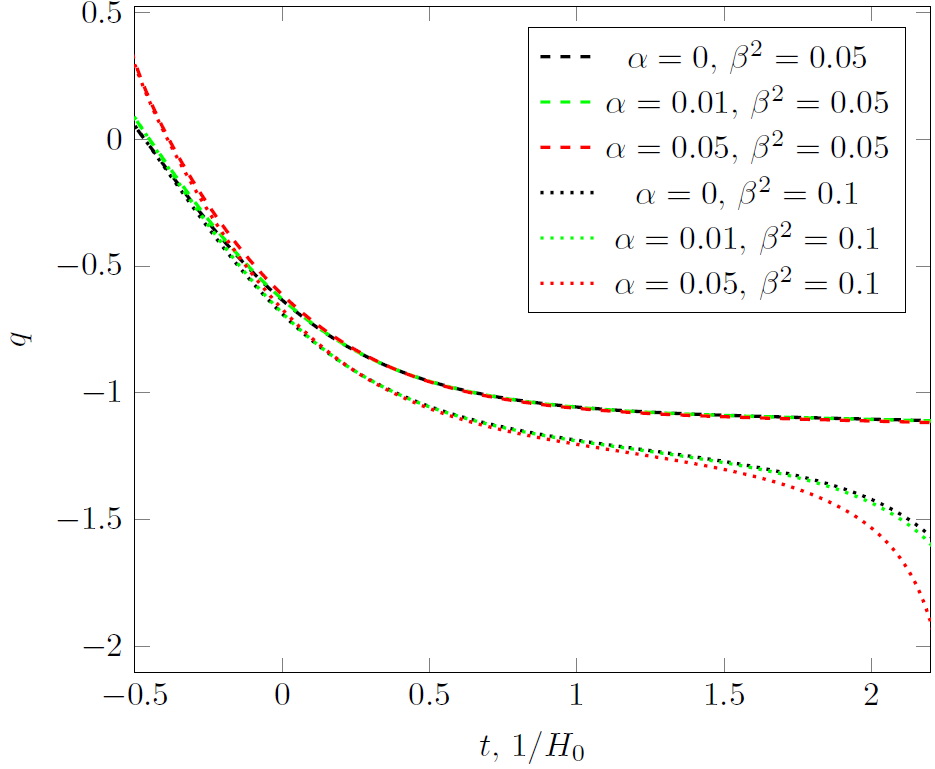}\includegraphics[scale=0.28]{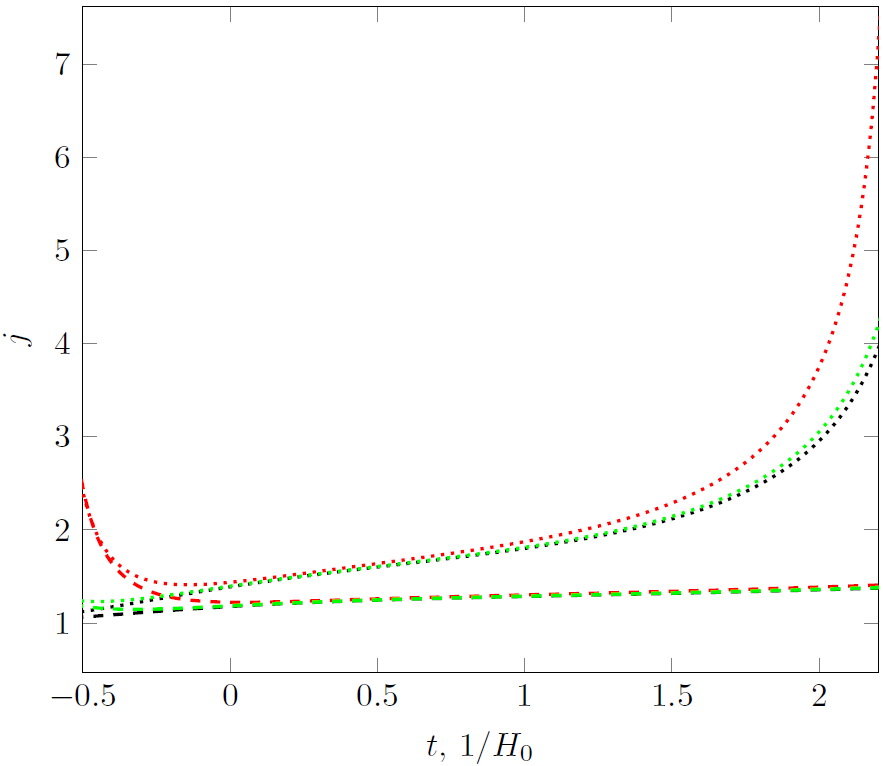}
    \caption{Deceleration (left panel) and jerk (right panel) parameters in
model ({\ref{EOS1}}) for $\gamma=2$ and some $\beta^2$ and
$\alpha$. Near the singularity state-finder parameters
$q\rightarrow -\infty$, $j\rightarrow\infty$ and its behaviour
significantly differs from model without thermal radiation.}
    \label{fig:2}
\end{figure}


\section{Pseudo-Rip with account of thermal radiation \label{SecIV}}

It is known from the literature that several phantom and
quintessence models can lead to the so-called pseudo-Rip
singularities \cite{Frampton-2,Astashenok}. The Universe expands
asymptotically according to exponential law with some effective
``cosmological constant''. It is interesting to investigate
influence of thermal radiation on the pseudo-rip expansion of this
sort. As an example of such model we shall consider the following,
\begin{equation}\label{EOS2}
    f(\rho_{d})=\pm \beta^2\rho_{d0}\frac{\rho_f-\rho_{d}}{\rho_f-\rho_{d0}}, \rho_f=\mbox{const}.
\end{equation}
We again choose the EoS so that the present time EoS parameter is
simply,
$$
w_{0}=\frac{p_{d0}}{\rho_{d0}}=-1\mp\beta^2.
$$
The choice of ``+'' ($\rho_f>\rho_{d0}$) corresponds to phantom
model, while the sign ``-'' ($\rho_f<\rho_{d0}$) describes
quintessence. For dark energy density as function of scale factor
we obtain,
\begin{equation}
    \rho_{d}=\rho_{f}-(\rho_f-\rho_{d0})a^{-\delta},\quad
\end{equation}
$$
\delta\equiv \frac{3\beta^2\rho_{d0}}{|\rho_f-\rho_{d0}|}.
$$
If $\rho_f<3/4\alpha$ then for large $a$, the energy density tends
to $\rho_f$, for phantom energy we have that $\rho_d\rightarrow
\rho_f-\epsilon$ while as for quintessence $\rho_d\rightarrow
\rho_f+\epsilon$.

The effective value of ''vacuum energy'' due to thermal radiation
is larger in comparison to $\rho_f$. For $\rho_f\rightarrow
\frac{3}{4\alpha}$ we have $\rho_{eff}/\rho_f\rightarrow 2$.
Therefore, the Universe expands faster due to the thermal
radiation. The dimensionless parameters $q$ and $h$ behave very
similar in both models {(see Fig. \ref{fig:3}).
\begin{figure}
    \centering
    \includegraphics[scale=0.28]{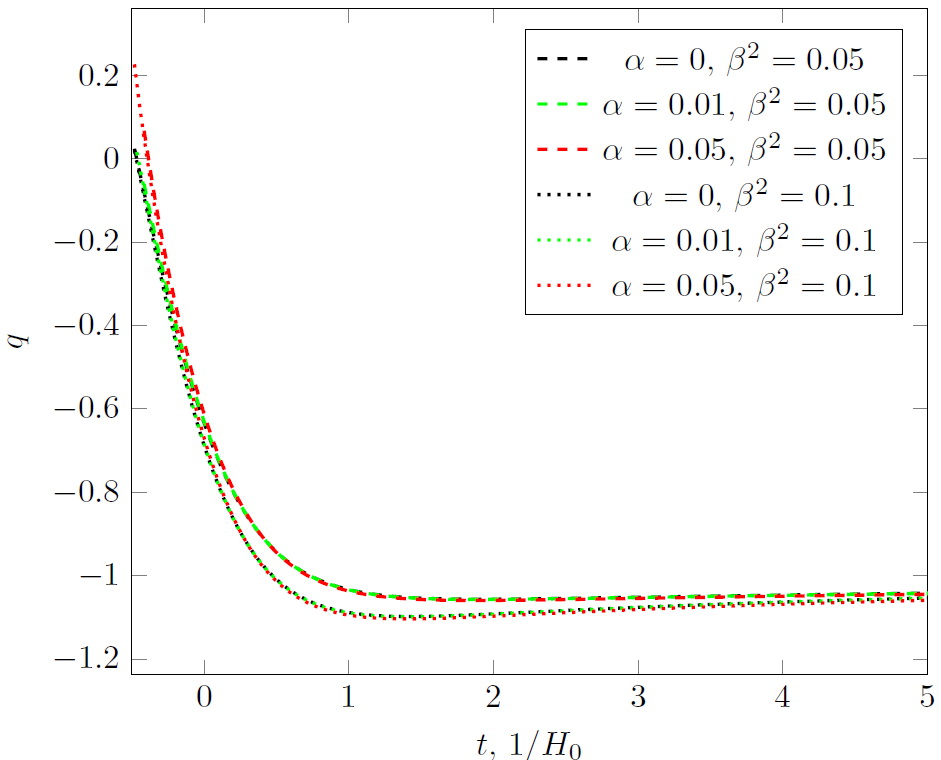}\includegraphics[scale=0.28]{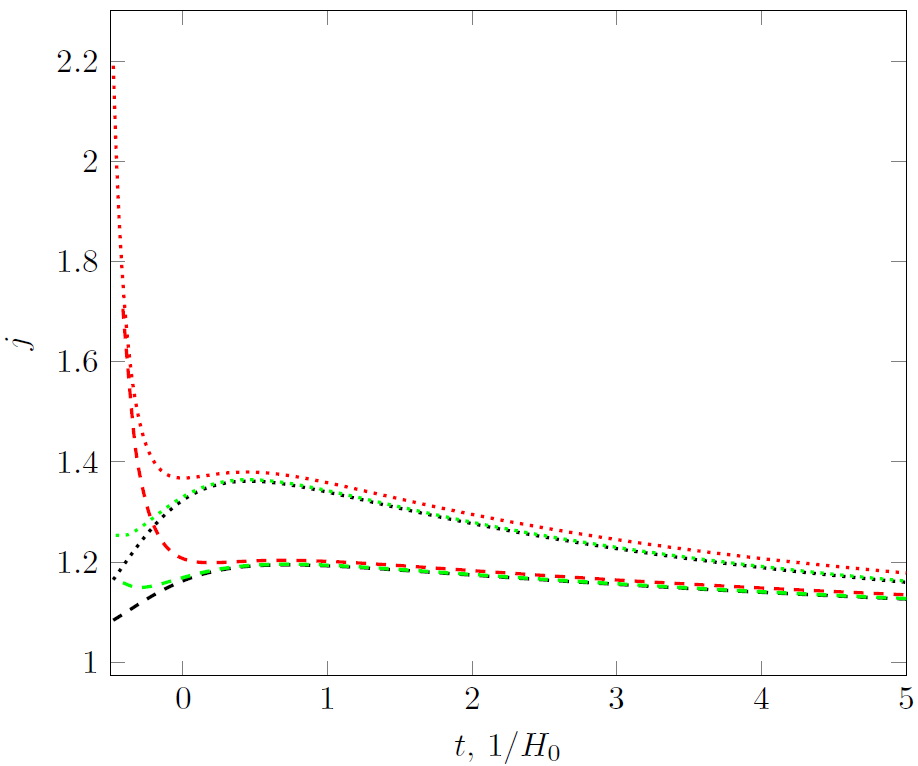}
    \caption{Deceleration (left panel) and jerk (right panel) parameters in model ({\ref{EOS2}}) for $\rho_f=20$ and some $\beta^2$ and $\alpha$.}
    \label{fig:3}
\end{figure}
For $\rho_f>3/4\alpha$, a sudden future singularity occurs before
exit on quasi-de Sitter expansion.

\section{Type II future singularity dark energy \label{SecVI}}

Some models of dark energy can lead to sudden singularity in
future without the effect of thermal radiation for example, \be
\label{TM}
f(\rho_{d})=\pm\beta^{2}\rho_{d0}\frac{1-\rho_{d0}/\rho_{f}}{1-\rho/\rho_{f}}\,
, \ee If we choose the sign ``+'', the energy density increases as
the Universe expands (phantom energy). The current EoS
parameter $w_{0}$ is simply \be w_{0}=-1-{\beta^2}. \ee
The pressure approaches
infinity and a sudden future singularity takes place. For
quintessence dark energy, $f(\rho_d)$ is negative and the final
singularity corresponds to $\ddot{a}<0$ (the so-called big crush). Singularity occurs earlier due to the thermal radiation (see Table III).

\begin{table}
\begin{tabular}{|c|c|c|c|c|c|c|}
\hline
    & \multicolumn{2}{c}{$w_{0}=-1.05$}\vline &
\multicolumn{2}{c}{$w_{0}=-1.075$}\vline &
\multicolumn{2}{c}{$w_{0}=-1.10$}\vline \\
\hline
$\tilde{\alpha}$ & $t_{f}-t_{0}$ & $t_{0}-t_{a}$ & $t_{f}-t_{0}$ & $t_{0}-t_{a}$ &
$t_{f}-t_{0}$ & $t_{0}-t_{a}$ \\
\hline
0       &    41.10  &   0.470    & 27.39 & 0.471  & 20.53 & 0.474 \\
0.001   &   40.99   & 0.467    & 27.31 & 0.468  & 20.47 &  0.470 \\
0.005   &   40.57   &  0.460    & 26.99 & 0.462  & 20.23 & 0.464 \\
0.01    &   39.81   &   0.452    & 26.52 & 0.454  & 19.88 & 0.456 \\
\hline
\end{tabular}
\caption{Time for singularity $t_f-t_{0}$ and time after moment
when acceleration began $t_{0}-t_{a}$ in model with sudden future
singularity (\ref{TM}) (in units of $H_{0}^{-1}$).  For initial
$\Omega_m$ the value 0.28 is taken. Parameter $\rho_f$ is 20 (in
units of $3H_{0}^2$). }
\end{table}
For phantom energy density, as function of scale factor, we have
the following relation,
\begin{equation}
\rho=\rho_{f}\left(1-\left((1-\Delta)^2-
6\beta^2 \Delta(1-\Delta)\ln a\right)^{1/2}\right), \quad
\Delta=\rho_{0}/\rho_{f}.
\end{equation}
The scale factor in past can be expressed as a function of the
redshift, which is defined as,
$$
a=\frac{1}{1+z}
$$
assuming that the present time scale factor is equal to unity, and
the dependence of the luminosity distance $D_{L}$ from the
redshift $z$ is \be \label{DL} D_{L} =
\frac{c}{H_{0}}(1+z)\int_{0}^{z}\left(\Omega_{m}
(1+z)^{3}+\Omega_\mathrm{D}h(z)\right)^{-1/2}d z, \ee
$$
h(z) =
\Delta^{-1}\left(1-\left(\left(1-\Delta\right)^2+6\beta^2\ln(1+z)\right)^{1
/2}\right).
$$
The model under discussion is indistinguishable from
the$\Lambda$CDM cosmology for $\Delta<<1$.
\begin{figure}
    \centering
    \includegraphics[scale=0.28]{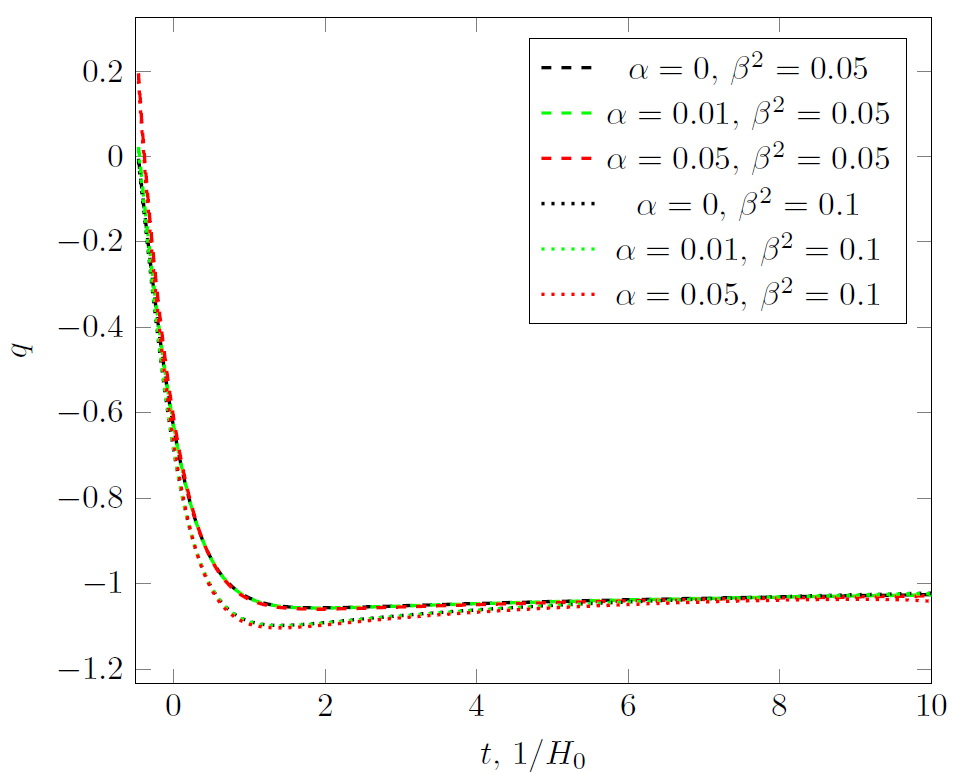}\includegraphics[scale=0.28]{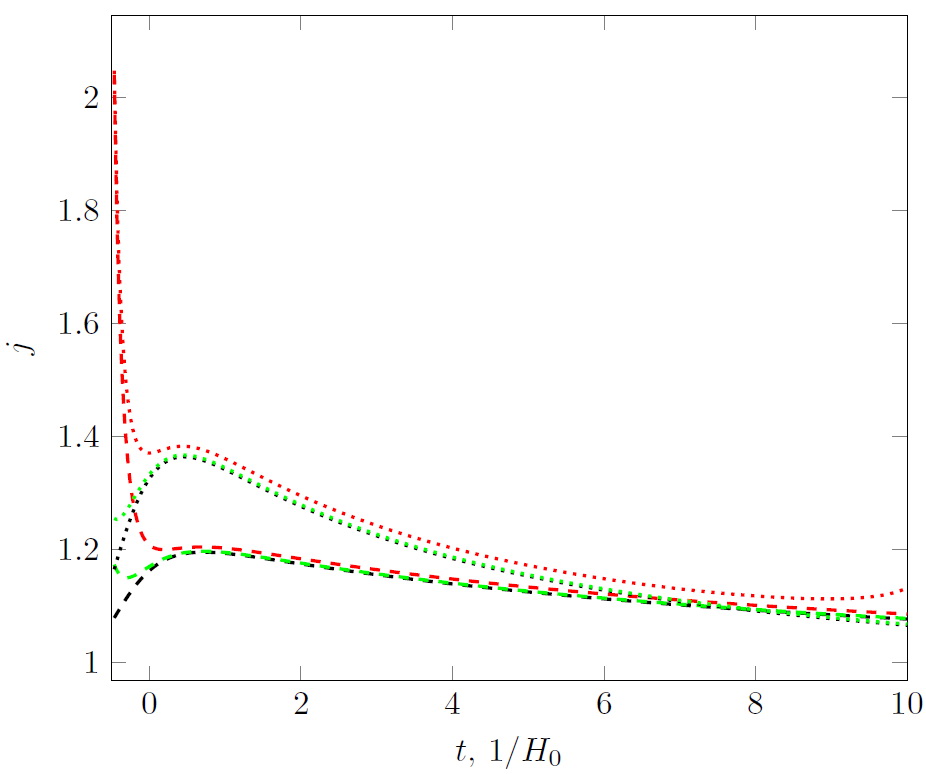}
    \caption{Deceleration (left panel) and jerk (right panel) parameters in model ({\ref{TM}}) for some $\beta^2$ and $\alpha$. Parameter $\rho_f$ is 20 (in units of $3H_{0}^2$).}
    \label{fig:4}
\end{figure}
{The deceleration and jerk parameters are given on Fig. \ref{fig:4}} for some
values of $\alpha$ and $\beta^2$. The example considered above is
a good theoretical illustration of dark energy models mimicking
vacuum energy, but these models lead to  Type II singularities.

\section{A Proposal for Evading the Big Rip Chaos: A Dark Fluid Mimicking $R^2$ Gravity}

In the process of approaching the Big Rip singularity, no matter
how this singularity occurs, the Hubble rate grows inevitably, and
a cosmological horizon of some sort is expected, as we already
discussed in previous sections. One intriguing theoretical
question is whether the singularity can be avoided in the first
place, if during the dark energy era some term appearing already
in the Lagrangian of the theory, makes the singularity milder, or
even disappear. In standard modified gravity contexts, such a
possibility is realized by adding $R^2$ terms in the Lagrangian,
see for example Ref. \cite{Bamba:2008ut}.

Thus it is tempting to add a geometric fluid with EoS parameter of
the form $w=w(H,\dot{H},\ddot{H})$ that mimics the $R^2$ gravity
energy density and pressure, and adding this near a Big Rip
singularity. Assuming a flat FRW spacetime, the Friedmann equation
near the Big Rip would be at leading order,
\begin{equation}\label{friedmaneqn}
\frac{3H^2}{\kappa^2}\simeq \alpha H^4+\rho_{\mathcal{G}}\, ,
\end{equation}
where $\rho_{\mathcal{G}}$ is the $R^2$ fluid energy density, the
analytic form of which as a function of the Hubble rate and its
derivatives with respect to the cosmic time is,
\begin{equation}\label{energydensityr2}
\rho_{\mathcal{G}}=-\frac{36 \beta  H \ddot{H}}{\kappa
^2}-\frac{108 \beta  H^2 \dot{H}}{\kappa ^2}+\frac{18 \beta
\dot{H}^2}{\kappa ^2}+\frac{3 H^2}{\kappa ^2}\, ,
\end{equation}
where $\beta$ is a parameter with mass dimensions $[m]^{-2}$.
Basically, the energy density $\rho_{\mathcal{G}}$ is essentially
the energy density corresponding to an $R^2$ geometric fluid of
the form $\beta R^2$, with the Ricci scalar being as usual
$R=12H^2+6\dot{H}$ for the flat FRW background. In order to see
whether the Big Rip singularity is avoided in the presence of the
$R^2$ geometric fluid, one must solve the Friedmann equation
(\ref{friedmaneqn}) analytically, which can be cast as follows,
\begin{equation}\label{freqnd}
\frac{36 \beta  H \ddot{H}}{\kappa ^2}+\frac{108 \beta H^2
\dot{H}}{\kappa ^2}-\frac{18 \beta  \dot{H}^2}{\kappa ^2}-3 \alpha
H^4\simeq 0\, .
\end{equation}
However, this is not an easy task, therefore we shall investigate
the behavior of the solutions numerically. We shall adopt the
reduced Planck units system, for convenience, in which
$\kappa^2=1$. The initial time instance of the time loop for our
numerical integration will be some initial point when the thermal
effects start to occur, so suppose that it is near,
$t\sim\mathcal{O}(1/H_{\mathcal{B}})$ where $H_{\mathcal{B}}$ is
the value of the Hubble rate when the initial Big Rip singularity
is approached, so it is expected to grow significantly near unity
in reduced Planck units, in contrast to the tiny present day value
in reduced Planck units. So assuming that we run the numerical
integration from $t\sim 0.$ in reduced Planck units, which is
basically at the beginning of the thermal effects, and by running
the numerical integration for $\mathcal{O}(10)$ time units, we
obtain the results presented in Fig. \ref{myplots}. Specifically,
in Fig. \ref{myplots} the behavior of the Hubble rate as a
function of the cosmic time in reduced Planck units is presented.
The time interval is significantly large in reduced Planck units,
so this numerical integration actually covers the a large time
interval in the far future after the thermal effects started to
have a measurable effect in the Friedmann equations. We used
various initial conditions for the derivative of the Hubble rate
in reduced Planck units, and and for all the plots, the Hubble
rate at the beginning of the time loop was assumed to be
$H_{\mathcal{B}}\simeq \mathcal{O}(10^{-1})$ and the values of
$\alpha$ and $\beta$ where assumed to be of the order of
$\alpha,\beta\sim\mathcal{O}(1)$. The red dashed curve corresponds
to $\dot{H}(0.)\sim \mathcal{O}(10^{-2})$, the red curve to
$\dot{H}(0.)\sim \mathcal{O}(10^{-1})$, while the blue dotted
curve to $\dot{H}(0.)\sim \mathcal{O}(10^{-5})$. As it is obvious
in all these cases, the Hubble rate grows significantly, but
instead of blowing up at finite-time, it reaches a plateau value,
which is different for the three different initial conditions, and
obviously this is a pure de Sitter state. This final de Sitter
approach of the Hubble rate is also common to pure $f(R)$ gravity
works, where the $R^2$ term eliminates completely the Big Rip
singularity from the cosmological evolution
\cite{Bamba:2008ut,Nojiri:2008fk,Capozziello:2009hc}. Clearly this
result indicates that the $R^2$ fluid stabilizes the cosmological
evolution, and definitely eliminates the Big Rip singularity.
Furthermore, as it was clearly shown in Refs.
\cite{Bamba:2008ut,Nojiri:2008fk,Capozziello:2009hc}, in the
presence of the $R^2$ term or other $f(R)$ gravity models, which
yield an exact de Sitter solution at future, apart from the Big
Rip singularity, it is also possible to eliminate Type II and III
singularities. Hence, the fluid mimicking the $R^2$ term in fact
eliminates not only Big Rip singularities, but also Type II and
Type III singularities, although in principle smooth types of
singularities, like the Type IV, can still occur. Furthermore,
eventually other fluids which lead to an accelerating Universe,
but yield asymptotically de Sitter solutions in the future, also
effectively cancel future singularities of Type I,II and III even
in presence of thermal effects.
\begin{figure}
    \centering
    \includegraphics[scale=0.5]{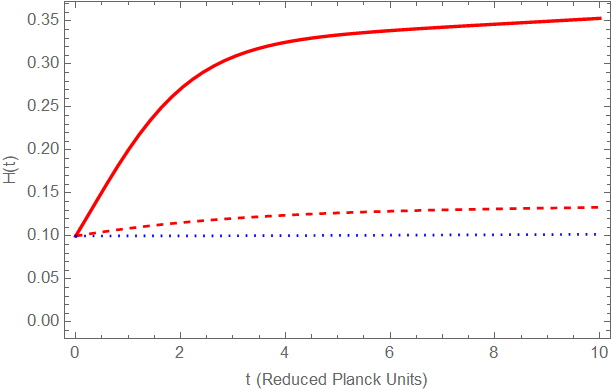}
    \caption{The Hubble rate as a function of the cosmic time in reduced Planck units, in the presence of an $R^2$ fluid. The final de Sitter point is reached
    and the Big Rip singularity is avoided, for various distinct initial conditions of $\dot{H}$, at the moment that the thermal effects affect the evolution.}
    \label{myplots}
\end{figure}

\section{Conclusion \label{SecVIII}}

{In this work we considered dark energy models by taking into
account thermal radiation effects. We assumed that this radiation
contributes to the total energy density with a energy density term
which is proportional to $H^4$, where $H$ is the Hubble rate.
Consequently, if the dark energy density values grow infinitely as
the Universe expands (phantom energy), then we can be certain that
a sudden future singularity takes place, or is approached
inevitably in the near future. For some finite energy density
$\rho_m$ and scale factor, the second derivative of scale factor
diverges. Thermal radiation leads to the so-called little rip
scenario, when the dark energy density increases infinitely with
time and the EoS parameter approaches the value $-1$
asymptotically at $t\rightarrow\infty$. The cosmological expansion
with a future Big Rip, without thermal radiation, also changes to
an evolution leading to a sudden singularity. It is interesting to
note that the phase of accelerated expansion in models with
thermal radiation, begins later in comparison with cosmological
models without this contribution, but the final singularity takes
place earlier.

The realization of a quasi-de Sitter expansion (pseudo-rip)
depends on the dark energy EoS. If the EoS parameter for dark
energy develops a zero value at some density $\rho_f$, then if
$\rho_f>\rho_m$, the sudden singularity still occurs. In the
opposite case the Universe evolves to a de Sitter regime with some
effective ``cosmological constant'', larger in comparison with
$\rho_f$. For $\rho_f$ close to $\rho_m$ we have
$\rho_{eff}/\rho_f\rightarrow 2$. In effect, the Universe expands
faster and evolves to the de Sitter expansion earlier. In the case
of an EoS with singularity (pressure tends to $\infty$ at some
finite $\rho_f$) we have in any case a sudden future singularity.
The inclusion of thermal radiation does not make worse the
compliance of the cosmological models with the observational data.
One can construct (as we have here) models that mimic the standard
$\Lambda$CDM model up to present time, but the contribution of the
thermal radiation can switch the future cosmological evolution to
a regime with a sudden future singularity.

At present time, dark energy remains a mystery and quite many
questions related to it are still not answered concretely. The
current models describing dark energy only answer some questions,
but to date no definitive answer is given. The main model which
remains with good compliance with the observational data is the
$\Lambda$CDM model, and thus many dark energy models originating
from various theoretical contexts, are mainly designed to mimic
the $\Lambda$CDM model. But the $\Lambda$CDM model has its weak
points, two of which already appear in its name, $\Lambda$ and
Cold Dark Matter. With regards to the latter, dark matter is a
speculation, but no dark matter particle has ever been observed.
With regard to $\Lambda$, the cosmological constant, its nature is
unknown, and if it is seen as the vacuum energy, its present day
value is extremely small and infinitely smaller from the predicted
vacuum energy value from quantum field theories. Apart from these
two, there exist other conceptual issues to be resolved in the
future, such as if the dark energy itself is dynamical or not, and
more importantly, is the dark energy era a phantom dark era? If
one sticks on the general relativistic approach and insist on
using he general relativistic recipe to describe the dark energy
era, so the usage of $\Lambda$ or scalar fields, the last two
questions will probably make their presence in a predominant way.
This is because in the case of dynamical dark energy with a
varying EoS, the cosmological constant would not fit at all, since
it yields a constant de Sitter EoS. Regarding the phantom
question, things are getting worse if one sticks with the general
relativistic recipe, scalar fields, since phantom scalar fields
must be used and phantom scalar fields from a theoretical point of
view are instabilities. Nevertheless, even if one adopts the
phantom scalar field description for the phantom dark energy era,
the result would be a Big Rip singularity as was demonstrated in
Ref. \cite{Caldwell-2003}. Thus the study of future finite-time
singularities is a possible eschatological scenario of our
Universe. In view of these questions, modified gravity in its
various forms, stands on a promising solid ground, since it
answers in a relatively successful way many of these questions. In
the models we presented in this work, we were able to generate
phantom evolution without the need for phantom scalar fields, and
we also indicated that finite-time singularities can actually
occur by using various fluid approaches, without again the need
for phantom fluids, like for example in general relativistic
contexts. In addition, some of the models can mimic at
present-time the $\Lambda$CDM model, and also can provide a viable
present day dark energy era, compatible with the Planck data. In
addition, we discussed how the evolution would change if we took
into account the thermal effects, thus our model indicates a road
map towards possible future evolutions. Of course our approach is
one of the many possible scenarios, however the future
observations are promising, thus as theorists we try to
investigate all the possible scenarios.

}

\section*{Acknowledgments}

This work was supported by MINECO (Spain), project
PID2019-104397GB-I00 and PHAROS COST Action (CA16214) (SDO).

\end{document}